# Building Reservoir Computing Hardware Using Low Energy-Barrier Magnetics

Samiran Ganguly* and Avik W. Ghosh
*Charles L. Brown Dept. of Electrical and Computer Engineering*
*University of Virginia*
*Charlottesville, VA 22904*
*\*sganguly@virginia.edu*

## ABSTRACT

Biologically inspired recurrent neural networks, such as reservoir computers are of interest in designing spatio-temporal data processors from a hardware point of view due to the simple learning scheme and deep connections to Kalman filters. In this work we discuss using in-depth simulation studies a way to construct hardware reservoir computers using an analog stochastic neuron cell built from a low energy-barrier magnet based magnetic tunnel junction and a few transistors. This allows us to implement a physical embodiment of the mathematical model of reservoir computers. Compact implementation of reservoir computers using such devices may enable building compact, energy-efficient signal processors for standalone or in-situ machine cognition in edge devices.

## 1 Introduction

Spatio-temporal inferencing is increasingly becoming an important technological necessity with proliferation of sensor-centric technology in society. Centralized "cloud" based processing of such sensor data, transmitted over a public data network, has been achieved with great success. However, latency, energy-efficiency, communication network reliability, and data privacy are longstanding concerns with this model of computing. This is raising the need for implementing AI algorithms in-situ with these sensors. There is a need to develop compact, light-weight, energy-efficient "edge" hardware computing fabric that performs a quick and dirty, but low latency inference locally that can be acted upon immediately, rather than depending on cloud support. This viewpoint is inspired by biological neural systems, where immediate stimuli are processed and acted upon by local neurons (edge hardware in this analogy) and only refined "world-model building" information is passed onto the brain (cloud in this analogy), allowing for ultra-efficient multi-scale spatio-temporal data processing capability spanning from hundreds of milliseconds to decades.

Application space for such in-situ machine cognition could be in unmanned vehicles, personalized health monitors, smart homes and appliances etc. that can keep working even in events of major disruption and unavailability of cloud computing support.

Reservoir computing [1] is a biologically inspired model of spatio-temporal inferencing that has been developed over last two decades [2]. This model of computation, developed in two distinct forms have certain features, to be discussed later, that make them attractive for hardware implementation. Indeed people have built such reservoir computers in hardware (see [3-10] for a few examples) using photonic devices, memristors, spintronic nano-oscillators, skyrmions etc.

In this work, we discuss an echo-state network (a particular model of reservoir computing) implementation using a previously proposed analog stochastic neuron device [11,12] built from a low energy-barrier magnetic tunnel junction in series with an n-channel MOSFET (metal oxide semiconductor field effect transistor), followed by an analog buffer device, such as a current mirror. This device produces a noisy sigmoidal neural activation function as its output voltage characteristics, in response to an analog voltage at its input node. Unlike other stochastic neuron proposals of similar designs [13-24] using a low energy-barrier magnetic tunnel junction with a binary response, this unit can be stabilized at any analog output voltage working as a true analog device, and is ideally suited to build an echo-state network which uses neurons with analog sigmoidal activations.

We show using circuit simulations the working principles of echo-state networks and two standard results widely used in the reservoir computing literature that demonstrates the feasibility of such a network working in practice.

The rest of the paper is organized as follows. In section 2 we provide an overview of the principles of reservoir computing, the structure, dynamics, and learning in such networks. In section 3 we discuss the connections between reservoir computing and extended Kalman filters which makes them useful in signal processing and control systems tasks. In section 3 we briefly discuss the physics of stochastic magnets and the analog stochastic neuron device built on this physics that allows for the desired noisy neuron characteristics in this unit. In section 5 we discuss the circuit-simulations of an echo-state network built from the analog stochastic neurons, illustrating the inner workings of the reservoir computing, as well as standard tasks popular in the community. We then conclude in section 6 with some additional comparison and thoughts on the implementation presented in this work with CMOS and a few other non-CMOS equivalents.

## 2 Reservoir Computing: A Short Introduction

Reservoir computing is a biologically inspired recurrent neural network model that has been used in processing of spatio-temporal data [25-29]. In this model of computing (Fig. 1 a), the input datastream (in time) is provided to a randomly interconnected collection of neurons. Each physical connection between two neurons introduces a certain delay to signal propagation from one neuron to the other. Also, the neurons are recurrently connected, which mean that a neuron can receive a delayed feedback response of its own activation after a few steps. The generalized response of neurons in this reservoir is given by:

$$x[t + \Delta t] = \alpha \tanh(W^{in}u[t + \Delta t] + W^{self}x[t] + W^{fb}y[t]) - \gamma x[t] + \beta v[t + \Delta t] \quad (1)$$

In this equation the vector $x[t]$ is the response of the neurons in the reservoir at time $t$, vector $u$ is the input signal, vector $y$ is the output of the reservoir computer (discussed below), $v$ is a noise term, the various weight matrices $W$ describe the interconnections: superscript *in* means input to the reservoir, superscript *self* means the interconnections with other neurons in the reservoir, superscript *fb* means the feedback connection with the *y*. The prefixes $\alpha, \beta, \gamma$ are the strength of a non-linear neural activation function (*tanh()* here), the strength of the noise signal, and the decay rate of signal at the neuron (considering leaky neurons) respectively. Note that this is the most generalized form of reservoir computing dynamics, and in literature people have used various forms omitting certain terms to choose a specific model that fits their particular computational needs, e.g. if stochasticity is not being considered, the noise term is dropped, if leakiness of neurons is not being considered the decay term is dropped, if feedback from the output is not considered, the $W^{fb}$ is dropped.

The reservoir computer works by sampling over the reservoir neural activations $x[t]$ to produce the output $y[t]$. This may be done either as a linear weighted sampling, or a deep neural network, or even another reservoir in form of a hierarchical reservoir architecture. In this work, we consider the simplest case, i.e. a linear weighted sampling given by:

$$y[t] = W^{out}x[t] \quad (2)$$

In this case, the learning in the network is performed by adjusting the sampling weights (i.e. the matrix $W^{out}$) to perform a desired inferencing task from the input $u[t]$. The standard approach is to use the Weiner-Hopf method of using the pseudo-inverse (*pinv()*) of a matrix. The input data $u[t]$ is imposed on the reservoir and the reservoir activation states $x[t]$ are collected to form a cumulative state matrix:$X[t, t + \Delta t, t + 2\Delta t, ... ]$, and the desired cumulative output matrix is formed as: $Y[t, t + \Delta t, t + 2\Delta t, ... ]$, the $W^{out}$ is then calculated as:

$$W^{out} = Y pinv(X) \quad (3)$$

It can be seen that the training is a one-shot process without any backpropagation through time (BPTT) which makes it an attractive option for fast online or evolutionary learning as the cumulative reservoir state matrix can be constructed on the fly and the weight update calculated continuously.

2.8 It should be noted that the version of reservoir computing presented above is called the "Echo-State Network" [30]. Another variant to reservoir computing is called the "Liquid State Machines" [31] where the neurons are spiking and the data encoding is in terms of spike intervals or density. The resulting spiking distribution in the reservoir can still be sampled in a similar fashion to generate an inference.

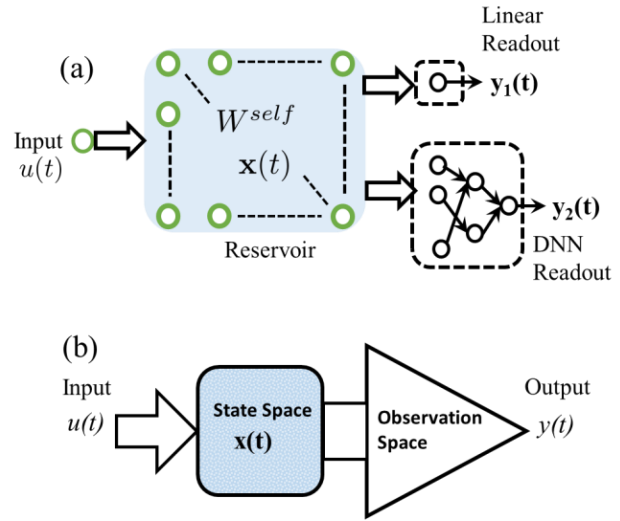

**Figure 1: Reservoir Computing and Extended Kalman Filters. (a) General schematic of a reservoir computer with a collection of recurrently connected non-linear neurons (connection topology and strength given by the matrix $W^{self}$). Analog sigmoid neurons are used in echo-state networks, while spiking neurons are used in liquid state machines. Multiple readouts may be attached to the same reservoir to extract multiple different inferences from the same signal, here we show a linear readout (used in this work) and a feed-forward deep neural network attached to the same reservoir. (b) General schematic of a Kalman filter, divided into two parts: a state space where a low dimensional signal is projected onto a high dimensional representation, and then an observation model extracts required information from this state space.**

## 3 Reservoir Computing: Connections to Extended Kalman Filters

The working principle behind reservoir computing is the ability to integrate and process various time slices of the input data $u[t]$, because the delay lines connecting the neurons cause the data to propagate through the network at various different speeds, therefore various different temporal samples of data, albeit transformed through neural activations $x[t]$ can be observed at any given moment by the output $y[t]$. The decay and noise adds a fading window to the maximum duration till which the oldest sample persists in the reservoir for sampling. This prevents the reservoir from going into chaotic states. However, highly damped reservoirs are incapable of processing the data to a "sufficient" depth in time. This property has been described in the literature as "computing at the edge of chaos", i.e. computing without stable states, and the corresponding dynamical behavior as "analog fading temporal memory", and the various temporal samples of the input data as "echo-states" [32].

From a signal processing point of view, the reservoir computing, in either avatars: echo-state networks or liquid state machines, closely resemble [33] the mathematical model of extended Kalman filters [34], an adaptive non-linear estimator that is widely used in a diverse set of areas such as control systems, navigation, time series data modeling, non-linear filtering, communications etc.

In Kalman filters, the signal is first provided to a dynamical system working as a state space, which is analogous to a reservoir of neurons, and the observation model samples from this state space (fig. 2). In Kalman filters the state space is linear, i.e. the equivalent neural activations are a linear function. However, the definition of Kalman filters have been expanded variously, and a Kalman filter with a non-linear activation function is called an extended Kalman filter. Mathematically extended Kalman filters are written as:

$$x_k = f_{NL}(x_{k-1}, u_{k-1}, k-1) + w_{k-1} \quad (4)$$

$$y_k = h(x_k, k) + v_k \quad (5)$$

Where eq.4 is the model for state space, while eq.5 is the observation model. The meanings of the symbols $u, x, y$ are the same as eq. and eq.2, $k$ denotes the discretized sample number ($t \to k, \Delta t \to 1$), while $w$ and $v$ are noise processes.

The working principle behind Kalman filtering, and indeed also reservoir computing is projection of low dimensional signal $u[t]$ onto a high dimensional state space through the dynamical phase space $x[t]$ of the Kalman filter state space/the reservoir of neurons, evident in the fact that the size of the state space/reservoir is much larger than the input data frame. This projection allows to separate and distinguish the features of the projected signal and their statistical properties with higher resolution. The observation model or the readout samples this high dimensional representation to form inferences. Multiple inferences may be made from the same higher dimensional representation as required. Attaching another reservoir hierarchically to the first one allows to project a specific high dimensional projected signal feature to even higher dimensionality.

Kalman filters' dynamical mathematical models are popularly implemented as a set of matrix-vector equations on a field programmable gate arrays (FPGAs) or application specific integrated circuits (ASICs). Reservoir computing, implemented on hardware can also be considered as an extended Kalman filter implementation and can be utilized for signal processing tasks.

In next two sections, we discuss a physical analog dynamical system implementation of reservoir computer and hence an extended Kalman filter leveraging the stochastic switching physics of low energy-barrier magnets.

## 4 Stochastic Magnetics, Stochastic Neurons

A magnet retains its state vector or magnetization direction due to its internal potential energy barrier ($U$) separating the two energy minima positions, up and down (Fig. 2a). $U$ is determined by the material properties and geometrical dimensions of the magnet and is given by $U = \frac{\gamma M_s H_k \Omega}{2}$, where $\gamma$ is the gyromagnetic ratio, $M_s$ is the saturation magnetization (material property), $H_k$ is the anisotropy energy field strength, determined largely by the magnet's shape, and $\Omega$ is the total volume of the magnet. For storage class memory elements $U$ is designed to be 40-60kT to provide a decade long state retention as such a high energy barrier is sufficient to prevent any thermal disturbance from stray magnetic fields. For a low $U$ magnet this state retention can be in 100s of *ps* to 10s of *ns* range.

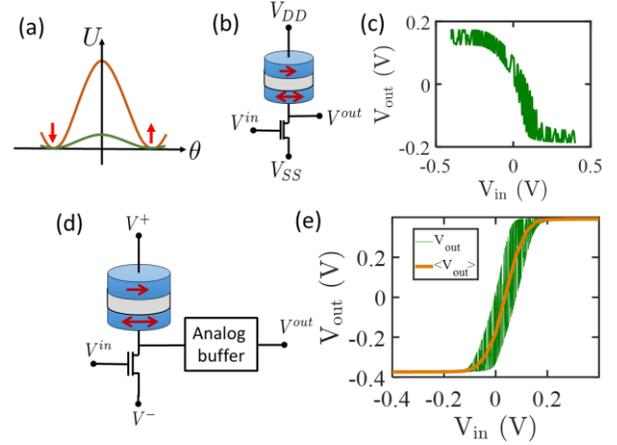

**Figure 2:** Stochastic Magnet based Analog Stochastic Neuron Device. (a) Energy landscape of a high energy-barrier and low energy-barrier magnet. (b) An embedded-MRAM unit with 1T-1MTJ structure. (c) Input-Output characteristics of a low energy-barrier magnet (LBM) based embedded-MRAM device. (d) Analog Stochastic Neuron (ASN) device from an LBM embedded-MRAM by cascading with an analog buffer. (e) Input-Output characteristics of the ASN device. The neuron's characteristics is noisy for low input signal levels (around 0V), while it saturates to a deterministic value at the high signal level (around ± 0.4V). The neuron can be stabilized at any intermediate values of the input voltage with the long term mean (yellow line in the plot) following a sigmoid or *tanh()* response.

In Fig. 2b we show the schematic of an embedded magnetic RAM (eMRAM) cell being used in the spin-transfer torque MRAM (STT MRAM) technology positioned as a non-volatile memory and a candidate for replacement of flash memory which is nearing its scaling limits. In this unit a magnetic tunnel junction is put in series with a selection nMOS transistor. Magnetic tunnel junctions are composed of two magnetic layers, separated by 1-2 *nm* thin MgO layer. One of these layers have very high energy barrier (150 kT or more), built either from thick (6-10 *nm*) magnet of a synthetic anti-ferromagnetic stack and is called the fixed layer. The other layer (1-3 *nm*) called the free layer is engineered to have the energy barriers we discussed above. The fixed layer's magnetization serves as a reference point for the cell whose resistance is minimum when the free layer's magnetization is parallel to the fixed layer, while it is maximum when it is anti-parallel to the fixed layer. The state of the cell can be read by applying a small voltage across it and detecting the current flowing through it using a sense amplifier. The cell may be written to a specific state or magnetization orientation of the free layer through self-generated spin current from the fixed layer which can "torque" the free layer to the appropriate direction (called the spin transfer torque effect). For a longer exposition on MTJ and STT MRAM technology please see [35, 36]. There are multiple commercial vendors who have now demonstrated capabilities for VLSI scale fabrication of STT MRAM technology with early commercialization.

We have earlier proposed [11,12] a modified version of this STT MRAM cell to build a stochastic neuron device. We and others have proposed (see references [13-24] for a short list of such proposals) using low barrier ($U$ = 1-5kT) magnets instead of U = 40-60kT magnets which turns the hysteretic transfer curve of such two-state memory cells (common with any non-volatile memory technology) into a transfer curve that looks like a *tanh()* as shown in Fig. 2c. However, this *tanh()* is only a long term sampling average of such a transfer curve and instantaneous response is much more noisy, and depending on the details of the design it can be "analog" like as shown or a "binary" curve where the instantaneous output voltage swing is from rail-to-rail but with still a long term sampling average of *tanh()*.

It is useful to add a buffer to this device to prevent loading and allowing us to use this in a circuit with impedance matching, gain and input-output isolation issues accounted for within the cell. In Fig. 2d we show such a structure with an analog buffer which can be made using standard circuits such as Wilson current mirror. We build a circuit model for this device using the Modular Approach to Spintronics [37] and Predictive Technology Model's [38] 14nm hp finfet transistor models and simulate in HSPICE. The output transfer characteristics (Fig. 2e) in this device turns into a true analog stochastic neuron's (ASN) activation, with a sigmoidal transfer curve and a white Gaussian noise enveloped on the sigmoid. It is possible to stabilize this cell's output at any point in the transfer characteristic. This unit therefore can work as a drop-in replacement for the mathematical model of an analog stochastic neuron in a circuit that physically implements a neural network with such neurons. The degree of stochasticity can be controlled to through internal cell design that we do not go in here but is presented elsewhere [11].

The output of this device, being a true sigmoid, can be used directly in a physical backpropagation algorithm implementation on hardware with well-defined derivatives, unlike with binary or spiking neuron's which do not have such well-defined derivatives and need either approximated derivatives or use other learning algorithms such as spike-time dependent plasticity based on Hebbian rule. The output characteristics of this device can be modeled as:

$$V_{out}(t) = \frac{V_{DD}}{2}\tanh(\beta V_{in}) + \alpha V_n(t, V_{in}) \quad (6)$$

In the above equation α, β are cell parameters (not to be confused with reservoir computer system parameters) discussed in [11], and $V_n$ is a Gaussian noise voltage whose instantaneous magnitude is time varying and its strength depends on the input voltage $V_{in}$.

Next we show how we can leverage the built-in analog stochastic neuron like behavior of this device to construct compact physical embodiment of an echo-state network variant of a reservoir computing.

## 5 Building Reservoir Computer with Stochastic Neurons

It can be seen from eqns. 1, 4 and 6 that the ASN device model includes within the physics and operation of the device certain important features necessary to implement reservoir computing/extended Kalman filters: a non-linear activation or transfer function, and noise. Therefore, this unit can be used as drop-in hardware unit to build a physical dynamical system that works as a reservoir computer/extended Kalman filter. Reservoir computing, however, incorporates another piece of physics which cannot be fully obtained from this device itself, viz. the signal decay.

The signal decay requires a discussion of the interconnects that connect together these neurons to build the reservoir. Physically the interconnects are RC delay transmission lines (for low frequencies). Therefore any signal transmitting through this delay line has a characteristic transfer or charging time from one neuron to the other. The ASN device itself does not have any built-in memory, being built out of all volatile components. Therefore as soon as the input signal turns off, the charged up device can "leak" the information away through various discharge paths in the circuit as well as within the device itself, which turns it into a leaky neuron which decays the signal/activation away. The interconnects can be designed in a way to provide a steady state signal decay of desired rate by balancing the charging and discharging resistive paths (see for [11] more detailed discussion). Therefore, it is possible to incorporate full functionality and the computing model of a reservoir computing in a physical circuit made out of ASNs and well-designed RC delay lines as interconnects. We can use controlled resistors, such as a linear mode biased transistor, or memristors, to provide the functionality of programmable reservoir connections or sampling network of the output to program a learned set of weights or reservoir topology as desired.

Building Reservoir Computing Hardware Using Low Energy-Barrier Magnetics

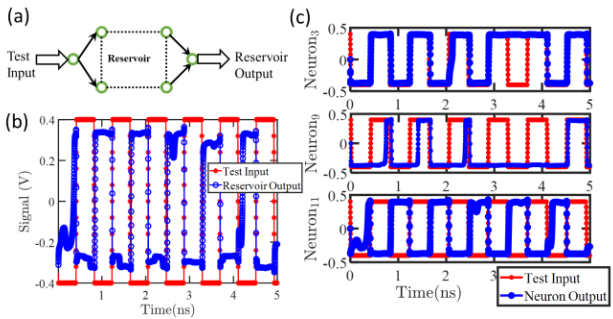

**Figure 3:** Reservoir Computing dynamics. (a) We setup a 25-node recurrently and randomly connected reservoir of analog stochastic neurons, with connections paths being implemented with delay lines of various time constants proportional to the connection matrix element strength. (b) The reservoir being tested to produce a signal inverter by a linear weighted sampling over the neurons in the reservoir. (c) activations of a few neurons in the reservoir in response to the input test signal as indicated in the figure. It can be seen that different neurons respond differently to the test signal, depending on the particulars of the network setup. In this case Neuron 11 produces the closest output to the desired functionality and is weighted more heavily in the sampling compared to neurons 3 and 9.

*It should be noted that we are proposing to physically build a reservoir computer by implementing a compact dynamical system composed of controlled low-barrier stochastic magnets, rather than emulating one numerically using a linear algebra accelerator.*

In Fig. 3 we show simulation results from a 25 node reservoir computer using a network of 25 ASN devices with RC delay lines for interconnects (fig. 3a), where RC delay is inversely proportional to the strength of the interconnection, i.e. a weaker connection means higher impedance, and larger delay interconnect line. The network was generated, trained, and tested in MATLAB. We then use auto-generated scripts to build SPICE netlists and ran on HSPICE 2016 using the *.trannoise* analysis which incorporates a Langevin like noise to transient circuit simulations which is well-suited for our simulation requirements. The simulation outputs were reimported back in MATLAB, processed, and plotted.

In Fig. 3b we show the input-output of the reservoir under a square pulse train. The network, in this case, was taught to reproduce the negative of its input signal, and hence the output observed in the plot. Even though the square pulse signal nominally looks "digital", it is treated in the simulation as an analog signal and the output is fully analog. We have tested this network using other pulse shapes (sinusoidal and triangular, not shown here) and it performs its tasks as intended. In these and other following simulations, the decay rate was 30% and noise magnitude was 5%.

In Fig. 3c we trace the co-evolutions of output of three of the neurons (labelled 3, 9, 11) of the reservoir with the test signal. It can be seen that the neurons respond differently to the input signal at any instance. For instance the neuron 3 follows the input signal for a few time steps, then picks up only the high value, while neuron 9 mostly follows the low value of the signal, only occasionally following the test signal (as blips), while neuron 11 runs as an inverse of the test signal. This exercise illustrates the central principle behind the idea of reservoir computing: a reservoir can sample over various aspects of the input signal over time, which allows for an inference to be made on the signal, the inference being a negator function generator in this case.

We now apply the network to two different signal processing tasks: (a) an adaptive video filter, (b) a temporal autoencoder. The process of setting up and running the network using MATLAB and HSPICE remains the same as before.

In Fig. 4 we demonstrate the use of a 200 node reservoir computer as a non-linear video filter capable of handling dynamic distortions and noise. We generate a synthetic video using a series of glyphs (two of them being "I" and "7" shown in the figure), which is used for training and testing. Keeping in mind the temporal nature of reservoir computing, the filtering is not performed on a frame-by-frame basis, in fact in our experience frame-by-frame filtering by reservoir computing is worse compared to a more traditional convolutional neural network based adaptive filter. Therefore, we provide the same glyph for multiple frames, however the distortion model is fast enough to change frame by frame. The non-linear estimation capability of the reservoir computer allows us to estimate and regenerate the true signal from such a severe distortion quite well. We have found 90-100% recovery rate in multiple tests on this network.

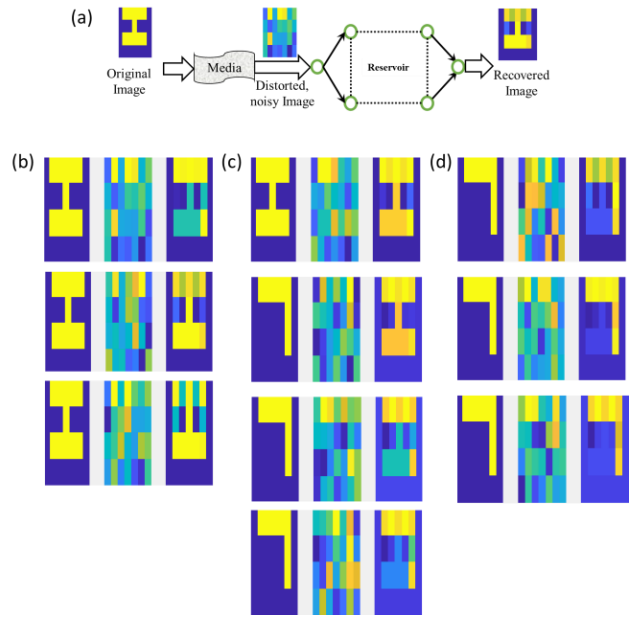

**Figure 4:** Reservoir Computer as a non-linear video filter. (a) video composed of character glyphs were provided to a non-linear noisy filter/transmission media that generates highly distorted noisy video. The reservoir has been trained to extract back the original signal from the distorted noisy signal. (b) A sequence of "I" glyphs from a few video frames (from top to bottom) being regenerated by the network shown as an image tuple: (left) original signal, (center) distorted signal, (right) recovered image. (c) A snapshot of frames when the video changes from "I" glyphs to "7" glyphs. The dynamical nature of the network is evident as it takes a few dataframes to transition to the "correct" glyph. (d) Sequence of "7" glyphs being regenerated back from noisy images.

Fig. 4a shows the scheme used for filtering. The media or the noisy non-linear distortion filter is a known function in this test scenario, in a "real world" application this can be unknown. This distorted video data is fed to the reservoir and the readout is trained to regenerate the original signal.

In Fig. 4b we show the result of filtering on a few frames of the video with the "I" glyph. The left image in these tuples is the original data, the center one is the distorted data obtained from the original data, and the right one is the recovered data. It can be seen that the network recovers the glyph to a high degree of accuracy in these examples. In Fig. 4d we perform similar filtering on the "7" glyph. In the Fig. 4c, we show a transition between the glyphs "I" to "7" and the resulting output. This example clearly demonstrates the dynamical nature of the filter as the network requires a few dataframes to "realize" that the underlying signal has changed to generate the correct image.

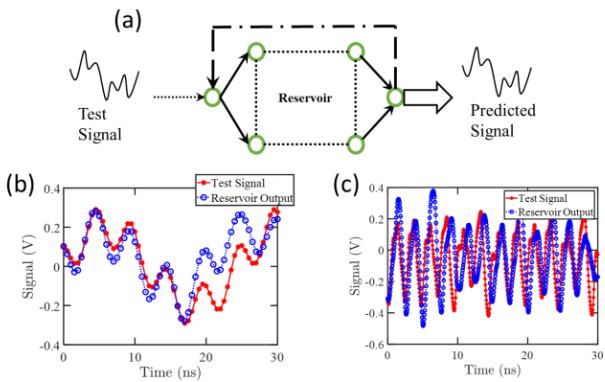

**Figure 5:** Reservoir computer as a sequence/ temporal autoencoder. (a) A test signal is provided to the reservoir to learn for a while. The test signal is then disconnected while the reservoir feedbacks to itself its own self-generated output which is then tested against the test signal. The reservoir learns the generative function of the test signal generator, therefore working as a multi-scale temporal autoencoder. (b) Test of the temporal autoencoder on a double sinusoid, i.e. two sinusoids of two different frequencies multiplied together. (c) Test of the temporal autoencoder on a Mackay-Glass time series equation.

In Fig. 5 we demonstrate the use of network as a sequence/temporal autoencoder. An autoencoder learns the underlying distribution or representation of a data model and can regenerate the signal on its own. A sequence/temporal autoencoder then learns the underlying generating function of a time series data or signal and can reproduce it in lieu of the original signal source. This is useful in navigation tasks such as trajectory prediction (also called non-linear autoregressive moving average or NARMA) which is a central application of Kalman filters and a widely used demonstration of reservoir computing in literature.

In Fig. 5a we first provide a test signal to the reservoir and teach it to reproduce the signal. We then decouple the test signal and feed the self-generated output of the network back to the input as shown. At this point the network is running without any external input and can generate the taught signal.

In Fig. 5b we show such an autoencoder on a signal generated from the product of two sinusoids of different frequencies. It can be seen that the network is capable of reproducing the signal running "blind", i.e. on its own and can do so with great accuracy for a while. After a while however the trajectory veers off due to inherent stochasticity in the network, and in such a case a corrective test signal injection can be used to "fix" the output. In Fig. 5c we perform the same task with a Mackay-Glass equation (MGE), which is a periodic but subtly chaotic sequence with a lot more information compared to the first sequence. It can be seen that the network's capability to reproduce this signal with accuracy is more limited compared to the first example since the dimensionality of MGE much higher. We have observed that larger networks reproduce the MGE to a better accuracy with the capability to follow each small nook and cranny of the signal.

These examples have appeared previously in reservoir computing literature (see [33] as an example) and have been chose specifically to show that a physical dynamical system built from ASN network can, in principle, demonstrate all these well-understood capabilities of reservoir computing and extended Kalman filters.

## 6 Discussion and Conclusion

In this work we demonstrated, using detailed circuit simulations, building of a dynamical systems based reservoir computer out of low energy-barrier magnetic analog stochastic neuron devices. Unlike the more usual CMOS based designs of neural network accelerators, we did not emulate or accelerate the mathematical model of the reservoir dynamics and sampling processes, rather the circuit operation of a network built from these neurons themselves gave rise to analog reservoir sates, while controlled interconnects formed signal delay lines that mix and match the various time samples of temporal data. Therefore, we claim that this method of implementing reservoir computers is much more compact than on a digital CMOS based alternative, due to massively reduced domain translation overheads, the domains being an analog stochastic dynamical system for ASN neurons, and a digital deterministic Turing Machine for a CMOS equivalent implementation. We have estimated elsewhere [11] the cost of pure digital, and mixed signal neurons implemented on FPGAs with the presented method. We see up to one to two orders of magnitude reduction in component count with ASN design with concomitant reduction in other metrics like lithographic area reduction and energy-delay product reduction.

In literature there are a multitude of non-CMOS approaches to building reservoir computer using a similar dynamical system implementation as in this work. Optoelectronics based systems are a popular choice (see [7] for a review). While photonic devices are

Building Reservoir Computing Hardware Using Low Energy-Barrier Magnetics

very tunable and can be made with high precision, they are much larger in size, resulting in much more energy consumption and building large scale circuits for computing are still a challenge compared to electronic devices. This makes the prospects of optoelectronics for edge computing hardware much more challenging compared to electronic, spintronic and memristive systems.

A multitude of proposals for reservoir computing using spintronic and memristive systems utilize various different aspects of these material systems, most typically the "state" information that can be stored or the special dynamics of these devices [8-10,18]. One such particular example [18] uses a low barrier MTJ based spin-torque nano oscillator (STNO). In this example a single oscillator state is sampled in a time multiplexed fashion. An external circuitry feeds back and processes these reservoir states back onto the oscillator as well as the readout. Therefore, this system utilizes the high degree of controllability of STNOs to develop reservoir states.

It is an extensive task to compare and contrast many such proposals with our approach. Indeed it may require an updated version of [39] to do so. Instead of taking that approach, we would briefly point to the extreme scalability of MTJs, inherently low-energy consumption, high controllability, extensive industry-wide capability of MTJ based VLSI fabrication, and strong research momentum of complementary and related low barrier MTJ based probabilistic computing efforts [20,40,41] as points that favor uptake of the presented unit for edge computing applications, as it presents an "evolutionary" rather than "revolutionary" change for the VLSI industry, both in terms of technology and economics.

A fair criticism laid at the doors of novel analog or dynamical systems based design are the problems of scalability and accuracy of computation from such systems, the prima facie reason for moving to all digital designs since 1980s. It is well known [42] that the analog systems require exponentially higher amount of energy to provide similar accuracy as an equivalent digital system, and the converse wisdom being that if low accuracy can be tolerated, analog systems can be exponentially low energy consuming compared to equivalent digital systems. The application space for a reservoir computing hardware as we envision is immediate signal processing and spatio-temporal inferencing on close-to-sensor-hardware edge devices, where power and compactness issues trump over very high accuracy. As an example, we can process a bio-physical signal, say ECG using an on-sensor hardware reservoir computer, which detects any anomalies such as arrhythmic beats within an ultra-low power chip, and in case of an inferred anomalous reading sends the data to a more capable cloud-based sophisticated neural network to determine accuracy of the diagnosis. Such a system can then implement a highly efficient cardiac monitoring system directly onto an ECG sensor that has a perceptible degree of in-situ machine cognition reducing its dependence on an external device, even if it is a body-area network smartphone for such services. Similar application spaces can be envisioned for other smart health applications, automated navigation, home automation, military applications etc.

## ACKNOWLEDGMENTS

This work was partially supported by the NSF I/UCRC on Multi-functional Integrated System Technology (MIST) Center IIP-1439644, IIP-1738752 and IIP-1439680.